\documentclass{appolb}
\usepackage{graphicx}
\usepackage{authblk}
\usepackage{amsmath}

\begin{document}
\title{Volume in the Extensive Thermodynamics of Black Holes: AdS and Kiselev spacetimes
\thanks{Presented in Prague, at VISEGRAD 2024 october}%
}
\author[1,3,4]{Réka Somogyfoki}
\author[2,3,4]{Péter Ván}

\affil[1]{ELTE Doctoral School of Physics, Eötvös Loránd University, Budapest, Hungary}
\affil[2]{Dep. of Energy Engineering, Faculty of Mechanical Engineering, Budapest University of Technology and Economics, Budapest, Hungary}
\affil[3]{Dep. of Theoretical Physics, Inst. of Particle and Nuclear Physics, HUN-REN Wigner Research Centre for Physics, Budapest, Hungary}
\affil[4]{Montavid Thermodynamic Research Group, Budapest, Hungary}

\maketitle
\begin{abstract}
Since black holes lack a straightforward notion of geometrical volume due to their event horizon structure and coordinate dependence, various approaches have been proposed to introduce a meaningful geometric and thermodynamic volume.  In this work we investigate the stability conditions of AdS black holes with and without volume.
\end{abstract}
 
\section{Introduction}

Thermodynamics of black holes violates some conventions of classical thermodynamics. Without volume, extensivity and thermodynamic stability of black holes as thermodynamic systems are particular. 

In AdS spacetimes one has the freedom to identify the mass of the black hole with either the internal energy or the enthalpy of a thermodynamic framework. In the second case, volume emerges from thermodynamics and stability is restored \cite{dolan2012pdv}. Also, the volumetric extension of the state space of simple black holes can lead to stability and extensivity, as was shown in \cite{biro2018black} using the ideas of thermodynamics of small systems by Hill~\cite{hill1994thermodynamics}.

The paper first compares some properties of Hawking--Page thermodynamics of AdS black holes with the volumetric extension of Dolan in \cite{dolan2012pdv}. Then explores how these principles may be applied by a more general pressure function and discovers a thermodynamic framework for Kiselev black holes, calculating the conditions of their stability.

\section{Hawking--Page phase boundaries}

\begin{figure}[htb]
\centerline{%
\includegraphics[width=6cm]{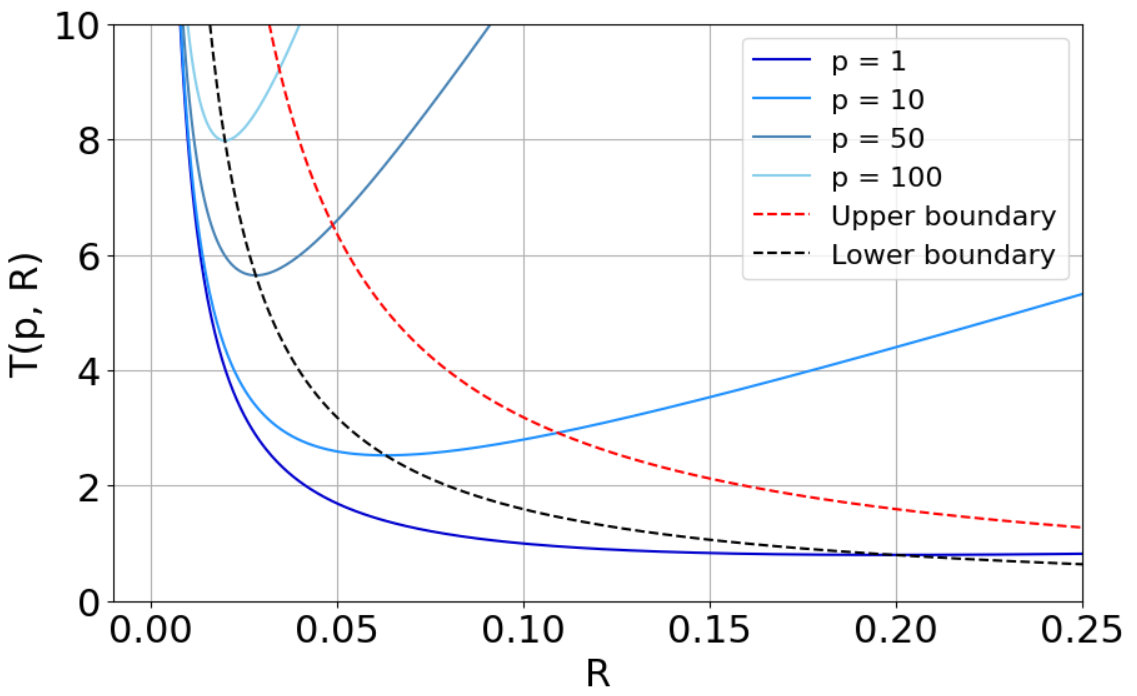}
\includegraphics[width=6cm]{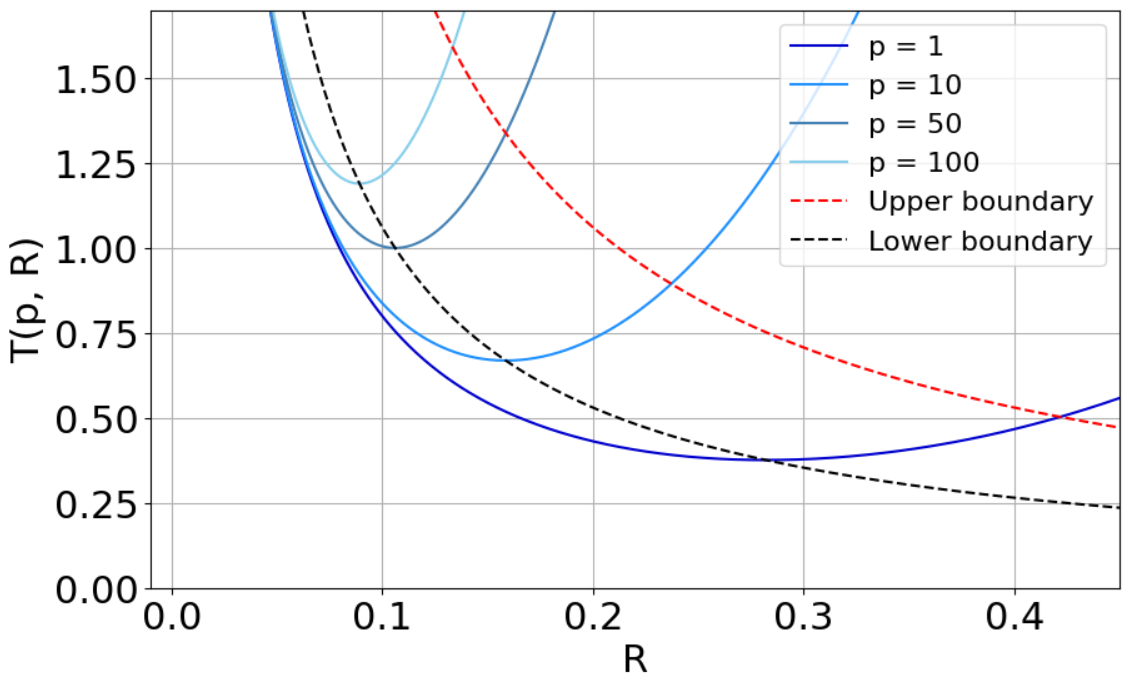}}
\caption{The dimensionless temperature $(T)$ expressed in terms of radius $(R)$ for AdS black holes with the Hawking--Page(--Dolan) stability and evaporation limits on the left and the same functions for Kiselev(--CR) black holes on the right.}
\label{Fig:THP}
\end{figure}

In thermodynamics of black holes in Anti-de Sitter space the geometrical and thermodynamical quantities are identified through the entropy definition as horizon area and through the horizon criterion 
\begin{equation}\label{thd_pots}
    M=\frac{R}{2}\left(1- \Lambda R^2\right), \qquad S(R) = \pi R^2,
\end{equation}
where the invariant mass $M=E(R,\Lambda)$ is the internal energy.
Here, $R$ is the radius of the horizon and $\Lambda$ is the cosmological constant. Then, the horizon- or Hawking temperature is identical to the thermodynamic one:
\begin{equation}
    T_{H}= \frac{d E}{d S} = \frac{1-3\Lambda R^2}{4\pi R}.
\end{equation}
Thermodynamic stability requires positive heat capacity and a further derivation gives the $-1/3\Lambda\leq R^2$ condition. Also, the AdS black hole is considered stable if its Helmholtz free energy $F(T) = E(T) - T S(T)$ is less than zero, therefore $1/\sqrt{-\Lambda}< R$ and the corresponding temperature must be greater than the marginal Hawking-Page temperature $T_{HP} =\sqrt{-\Lambda}/\pi$. Above $T_{HP}$, the system favors black hole formation over pure thermal radiation and black holes remain stable. Below $T_{HP}$ their free energy becomes positive, causing evaporation. Left side of Fig. \ref{Fig:THP} shows the temperature-radius functions with various cosmological constants and also the boundaries of thermodynamic and evaporation stability. 

\section{Thermodynamic volume following Dolan}

Alternatively, the thermodynamic state space of AdS black holes can be extended by volume, treating the black hole mass as enthalpy and the cosmological constant as pressure, more precisely $p = -3\Lambda/8\pi$. This way, we are in AdS spacetime ($p \geq 0$) but $\Lambda$ is not a constant anymore \cite{KasEta09a,dolan2012pdv}. 
\begin{equation}\label{H}
    M=\frac{R}{2}\left(1- \Lambda R^2\right)=H(S,p)=\frac{1}{2}\sqrt{\frac{S}{\pi}}+\frac{4\pi}{3}p\Big(\frac{S}{\pi}\Big)^{3/2},
\end{equation}
where the radius is expressed with the entropy defined in \eqref{thd_pots}. 
The partial derivatives of the enthalpy give us the thermodynamic temperature which is identical with the Hawking temperature, and the black hole volume:
\begin{equation}
    \left.\frac{\partial H}{\partial S}\right|_p = T =\frac{1-3\Lambda R^2}{4\pi R}, ~~~~~~~~~~~~
    \left.\frac{\partial H}{\partial p}\right|_S =V=\frac{4\pi}{3}R^3.
\end{equation}
Thermodynamic stability criteria emerge from nonnegative izochoric heat capacity $\left.\frac{\partial E}{\partial T}\right|_V \geq 0 $, and nonnegative isothermal compressibility $\left.\frac{\partial p}{\partial V}\right|_T \leq 0 $. The internal energy is $E = H - pV = R/2$, which makes the first condition result in $C_v = 0$. The second condition $R\geq (2 \pi T)^{-1}$, gives the minimum of $T(R)$. This condition is identical to the one we have obtained with the original Hawking--Page thermodynamics from the condition of nonnegative heat capacity.  

If the Gibbs potential $G = H - TS$ is below zero, the thermodynamic system does not emit particles to a radiation field\footnote{Which has zero Gibbs potential.}. One can best calculate this as the function of the pressure and the radius of the black hole, and the condition $R\geq 1/\pi T$ emerges, which gives the Hawking-Page temperature in the zero evaporation limit\footnote{Let us mention, that there is no Maxwell-rule at the background \cite{spallucci2013maxwell}, because here the Gibbs potential is not related to a chemical potential, lacking a particle exchange.}. The left image in Fig. \ref{Fig:THP} illustrates both the radius dependence of the temperature for Dolan's thermodynamics as well as for the Hawking--Page one. 

\section{Volume and extensivity}

The previous thermodynamic systems are degenerate. The Hawking--Page one is characterized by a single thermodynamic state variable, the internal energy (or entropy). Its Gibbs potential does not exist. Also, in the second system the volume and entropy are not independent: $S$ and $V$ mutually determine each other: $V$ looks redundant. 
The degeneracy is inherited from Schwarzschild thermodynamics, and it can be resolved assuming that the radius encodes a scaling property. Here the scaling $R(E,V)=R_S E^\alpha V^\beta$ is chosen according to \cite{biro2018black, biro2020volume}. Several thermodynamic quantities can be calculated without further ado, e.g. the pressure is $p=\beta E/\alpha V$. 

We have seen that the internal energy is $E=R/2$ when the invariant mass is identified with the enthalpy. The entropy can be expressed as
\begin{equation}
    S\big(R(E,V)\big)=\pi R^2 \approx E^{2\alpha}(R^3)^{2\beta},
\end{equation}
so $2 = \alpha+3\beta$ must stand. If the pressure corresponds to black body radiation, meaning $\alpha =3 \beta$, then $\alpha= 1/2$ and $\beta =1/6$. Hawking temperature is identical with the thermodynamic temperature again. Moreover, one can generalize the thermodynamical framework, where the pressure depends on both the internal energy and the volume. As an example, let us consider the following general interpretation of the same horizon condition:
\begin{equation}\label{H3}
    M=\frac{R}{2}\left(1- \Lambda(R,p) R^2\right)=H(S,p)=\frac{1}{2}\sqrt{\frac{S}{\pi}}+\frac{16\pi^2}{15}p\Big(\frac{S}{\pi}\Big)^{5/2},
\end{equation}
where $\Lambda = -2 p (2\pi R)^2/15$ is pressure and radius dependent. The volume
\begin{equation}
    V=\frac{\partial H}{\partial p}\Big|_S=\Big(\frac{S}{\pi}\Big)^{5/2}=\frac{(4 \pi)^2}{15}R^5
\end{equation}
is consistent with the Christodoulou–-Rovelli volume \cite{christodoulou2015big}, namely, that it is proportional to $R^5$. Then the stability and evaporation conditions become $1/3\pi T \leq R\leq 2/3\pi T$. The corresponding temperature-radius functions are shown in the right image of Fig. \ref{Fig:THP}.

From the radius depence one can recognise that this is a Kiselev black hole \cite{Kis03a} with parameters $w=-5/3$ and $K=-(4 \pi)^2/15$
. Remarkably, scaling relations are exactly fixed, since the average pressure $\bar{p} = w E/V$ together with extensive, first order Euler homogeneous entropy where $\alpha+\beta =1/2$ requires $\alpha = -3/4$ and $\beta = 5/4$.  

Black holes in Kiselev spacetimes are in inhomogeneous fluid-field environment, with a diagonal energy momentum $T^{ab} = diag(\rho, p_r, p_t, p_t)$ and the relations $\rho = -p_r$ $p_t = 2 p_r$ between the energy density and pressure terms \cite{Vis20a,CziIgu25a}. The fluid is isotropic only if $w=-1$, then it gives the AdS black hole back. We can see this value is inconsistent with Dolan's thermodynamics, but can be consistently interpreted with the scaling assumption with $\alpha=-1/2$ and $\beta = 1/2$. This way $\alpha +\beta \neq 1/2$, the black hole is not an extensive thermodynamic body.

\section{Discussion}

In the general Kiselev case, the boundary conditions of the fluid–field environment may appear artificial. However, considering the thermodynamic stability of the system, it is reasonable to expect that the spacetime singularity, a black hole, transforms its continuous environment rather than existing within a fixed one. This perspective follows naturally from the dynamical system framework of ordinary thermodynamics \cite{matolcsi2004ordinary,Had19b}.

Average pressure may be a too simple thermodynamic representation. Our calculations indicate that mixed fluid-radiation environment with anisotropic energy momentum requires the extension of the thermodynamic framework.

\section{Acknowledgement}   
The work was supported by the grant NKFIH NKKP-Advanced 150038.

\bibliographystyle{unsrt}

\end{document}